\documentclass[11pt,dvips,twoside,letterpaper]{article}
\usepackage{graphicx}
\usepackage{pslatex}
\usepackage{fancyhdr}
\usepackage{geometry}

\usepackage[T1]{fontenc}
\usepackage[latin1]{inputenc}
\usepackage{amssymb}

\usepackage{graphicx}
\usepackage{epsfig}

\begin{document}
\title{
~\\[-1.2in]
{\normalsize\noindent
\begin{picture}(0,0)(228,-3.15)
\begin{tabular}{l p{0in} r}
In: New Developments in Ferromagnetism Research  & & ISBN 1-59454-461-1 \\
Editor: V.N. Murray, pp. 101-120  & & \copyright 2005 Nova Science Publishers, Inc.\\
\end{tabular}
\end{picture}}
{\begin{flushleft} ~\\[0.63in]{\normalsize\bfseries\textit{Chapter~4}} \end{flushleft} ~\\[0.13in] 
\bfseries\scshape Ferromagnetic Domain Walls in finite\\ systems: mean-field critical exponents\\
and applications}}
\author{
\bfseries\itshape  B. Uchoa and G. G. Cabrera\\ Instituto de F\'{\i}sica {}``{ Gleb Wataghin}{ '',}
Universidade Estadual\\ de Campinas (UNICAMP),
  C. P. 6165, Campinas, SP 13083-970, Brazil}
\date{}
\maketitle
\thispagestyle{empty}
\setcounter{page}{101}

\pagestyle{fancy}
\fancyhead{}
\fancyhead[EC]{B. Uchoa and G. G. Cabrera}
\fancyhead[EL,OR]{\thepage}
\fancyhead[OC]{Ferromagnetic Domain walls in finite systems: mean-field critical exponents ...}
\fancyfoot{}
\renewcommand\headrulewidth{0.5pt}
\addtolength{\headheight}{2pt} 
\headsep=9pt


\begin{abstract} 
The distribution of magnetic moments in finite
ferromagnetic bodies was first investigated by Landau and Lifshitz
in a famous paper \textit{\emph{{[}}}\textit{Phys. Z. Soviet Union},
\textbf{8}, 153 (1935){]}, where they obtained the domain structure
of a ferromagnetic crystal at low temperatures, in the regime of saturated
magnetization. In this article, we investigate the general properties
of ferromagnetic domain walls of uniaxial crystals from the view point
of the Landau free energy. We present the basic ideas at
an introductory level, for non-experts. Extending the formalism to
the vicinity of the Curie temperature, where a general qualitative description
by the Landau theory of phase transitions can be applied, we find
that domain walls tend to suppress the layers, leading to a continuous
vanishing of the domain structure with anomalous critical exponents.
In the saturated regime, we discuss the role of domain walls in mesoscopic
systems and ferromagnetic nanojunctions, relating the observed magnetoresistance
with promising applications in the recent area of spintronics.
\end{abstract}

\section{Introduction}

The distribution of the magnetization inside a general ferromagnetic
body follows a closed flux configuration which leads to the appearance
of magnetic domains. For the stripe domain structure, which is common
in whiskers, magnetizations of neighboring domains are oppositely
oriented, separated by $180^{\circ}$ Bloch walls of finite width.
Inside the domain walls, the change of the magnetization is not discrete
but smooth. In a pioneering work, Landau and Lifshitz proposed for
the first time a theory that quantitatively predicted the above configuration,
relating domain sizes and wall width with the dimensions of the body
and some phenomenological parameters associated with the crystal structure\cite{ld}.
In their analysis, the magnetic energy of the crystal is built as
consisting of two terms: one is the exchange interaction of the spins,
proportional to \begin{equation}
\lbrack(\nabla m_{x})^{2}+(\nabla m_{y})^{2}+(\nabla m_{z})^{2}]\,\,,\label{exchange}\end{equation}
 where $\mathbf{m=(}m_{x},m_{y,}m_{z})$ is the magnetic moment, whose
absolute value is considered constant and equal to the saturation
value; the other gives the contribution of the magnetic anisotropy
of the crystal, that competes with the exchange interaction. Assuming
an easy direction of magnetization along the $z$-axis, the latter
was written as \begin{equation}
\beta[m_{x}^{2}+m_{y}^{2}],\label{anisotropyTerm}\end{equation}
 with $\beta>0$. The exchange term (\ref{exchange}) considered by
Landau and Lifshitz can be obtained from the classical Heisenberg
model,\[
H=-J\sum_{i}^{N}\sum_{\delta}\,\mathbf{S}_{i}\cdot\mathbf{S}_{i+\delta}\,,\]
 (with $J>0$, and $\delta$ indexing the nearest neighbor sites)
in the continuous approximation, replacing discrete spin variables
$\mathbf{S}_{i}$ by a spin density order parameter (magnetization
density) $\mathbf{m}(\mathbf{r})$,
spatially averaged in a given local configuration of spins \cite{kittel}.

To find the distribution of the magnetization inside the material,
one solves a variational equation that minimizes the sum of these
two contributions for the particular geometry and size of the sample,
with proper boundary conditions. For the stripe geometry and the closed
flux configuration, boundary conditions induce the formation of stripe
domains. The solution of the problem leads to a soliton-like pattern
for the magnetization near domain walls\cite{ld}, forming a non-homogeneous
phase where the order parameter (magnetization) changes sign alternately
from one domain to the next. Due to the ferromagnetic interactions,
the spins locally tend to be aligned, and one sees that the exchange
energy is small everywhere, except in the intermediate region between
domains, where the magnetic moment smoothly changes its orientation
to satisfy the anisotropic energy and the boundary conditions. The
overall energy is minimized, since there are no field lines outside
the sample. The main assumption of this treatment is that the system
is ferromagnetically ordered and close to saturation, in the low temperature
phase. An extensive description of the general properties of magnetic
domains and domain walls is available in Ref. \cite{hubert}.

There are several important applications related to the existence
of magnetically ordered domain walls, including the recent advances
in spintronics for electronic devices. The purpose of this article
is to discuss the structure of the domain walls in a phenomenological
way, using the simple but efficient tools provided by the Landau analysis
on ferromagnetism. In the first part of this article we consider
the mean field critical regime of domains walls, while the second
one is devoted to some effects in the saturated
regime, where we discuss possible applications.

To understand how the domain structure behaves in the vicinity of
the Curie temperature, where the magnetic order becomes paramagnetic,
we extend the original Landau-Lifshitz approach within 
the more general framework of the 
Landau theory of phase transitions. The \emph{Landau free energy}
of a ferromagnetic system is written as a series of the magnetic moment
density $\mathbf{m}=(m_{x},m_{y},m_{z})$, which is considered as
the order parameter (OP). In the neighborhood of the critical temperature
$T_{c}$, $|\mathbf{m}|$ is assumed to be small. This power series
introduces phenomenological coefficients for the exchange and anisotropy
contributions, and is written following the general Landau prescription.
If one writes the system degrees of freedom in terms of the magnetization
density $\mathbf{m}$ and the magnetic field $\mathbf{H}$, the thermodynamics
is obtained through the partition function \begin{equation}
Z=\exp\!\left[\frac{-G(\mathbf{H},T)}{k_{B}T}\right]\,,\label{Z}\end{equation}
 which is written in terms of the Gibbs free energy $G(\mathbf{H},T)$,
and is proportional to \begin{equation}
\int(D\mathbf{m})\exp\!\left[-\frac{E(\mathbf{m},\mathbf{H})}{k_{B}T}\right]\,,\label{partition}\end{equation}
 where $k_{B}$ is the Boltzmann constant and $T$ is the temperature.
The symbol $D\mathbf{m}$ means integration over all possible configurations
of $\mathbf{m}$, with $E(\mathbf{m},\mathbf{H})$ being the effective
Hamiltonian

\begin{equation}
E(\mathbf{m},\mathbf{H})=\int d^{d}r\,\, f(\mathbf{m},\mathbf{H})\,.\label{action}\end{equation}
 and $f$ the Landau functional. The latter integration is carried
for a $d$-dimensional system. The expression (\ref{partition}) yields
the partition function as a functional integral of the field. Assuming
that the dimensions of the stripe domains are large in comparison
with the lattice constant (continuous approximation), variations of
the magnetization lines inside the material are considered smooth
and the term that measures the inhomogeneity contribution can be treated
in first order as a gradient. In other words, we assume that the field
$\mathbf{m}=(m_{x},m_{y},m_{z})$ and its derivatives are continuous.
In the mean field level, the minimization of the Gibbs free energy
given by Eq. (\ref{Z}) is done in the saddle point approximation,
\[
\left.\frac{\delta E(\mathbf{m},\mathbf{H})}{\delta\mathbf{m}}\right|_{\mathbf{m}=\langle\mathbf{m}\rangle}=0\,,\]
 with the symbol $\delta$ representing the variation of  integral
(\ref{action}) with respect to the magnetization, calculated at the
saddle point $\mathbf{m}=\langle\mathbf{m}\rangle$, where the brackets
have the meaning of an ensemble average. In this approximation (which
neglects the effects of fluctuations around the saddle point), the
Landau functional $f(\mathbf{m},\mathbf{H})$ plays the role of the
free energy density and is the relevant quantity to be minimized.
In the vicinity of the phase transition, however, the fluctuations
around the saddle point $\delta\mathbf{m}=\mathbf{m}-\langle\mathbf{m}\rangle$
acquire a major importance for physicals systems of low dimension and 
coordination
number. Defining $\xi$ as the correlation length, we notice that the
mean field analysis breaks down when the amplitude of the fluctuations
measured by the correlation length $\langle\delta\mathbf{m}(\mathbf{r})\delta\mathbf{m}(\mathbf{0})\rangle\propto\xi^{2-d}$
is larger than the square of the mean field order parameter $\langle\mathbf{m}\rangle^{2}\propto\xi^{-2}$
\cite{chaikin}. Since $\xi$ diverges in the critical point, the
saddle point approximation excludes the region immediately bellow
$T_{c}$ for $d<4$, where the fluctuations predominate.

\subsection{Phenomenological Landau Theory}

It is instructive to develop a few basic ideas regarding the general
theory of phase transitions in a continuous medium. The free energy
$f$ is built under symmetry considerations, observing the most general
form of $f$ which is preserved under all the symmetry operations
of the OP $\mathbf{m}$ that leave the system physically invariant.
As we are in the continuum, we only need to know the point group symmetry
of the OP, which for a ferromagnet depends basically on the anisotropy
directions in the crystal. In general grounds, the free energy is
made of three terms. The first one is a series expansion in the OP
derivatives $K(m_{i},\frac{\partial m_{i}}{\partial r_{j}})$, which
we call ``kinetic'' term,\[
K\equiv\sum_{ij}A_{ij}\frac{\partial m_{i}}{\partial r_{j}}+\sum_{ijk}B_{ijk}m_{k}\frac{\partial m_{i}}{\partial r_{j}}+\sum_{ijkl}C_{ijkl}\frac{\partial m_{i}}{\partial r_{j}}\frac{\partial m_{k}}{\partial r_{l}}\,,\]
 where $r_{i}$ are the space coordinates and $m_{i}$ are the OP
components. Not all of these terms are to be kept, since first order
derivatives on the form $\sum_{ij}\int_{V}\textrm{d}^{d}r\,\frac{\partial m_{i}}{\partial r_{j}}$
are non-extensive when integrated in the volume $V$. In the same
way, part of the second term above \[
\sum_{ijk}\frac{1}{2}\left(B_{ijk}+B_{kji}\right)\left[m_{k}\frac{\partial m_{i}}{\partial r_{j}}+m_{i}\frac{\partial m_{k}}{\partial r_{j}}\right]=\sum_{ijk}\frac{1}{2}\left(B_{ijk}+B_{kji}\right)\frac{\partial}{\partial r_{j}}\left[m_{k}m_{i}\right]\]
 is also non-extensive in the volume. Keeping only the extensive terms,
the most general form of $K$ up to second order is \cite{tandt}\begin{equation}
K\equiv\sum_{ijk}\frac{1}{2}\left(B_{ijk}-B_{kji}\right)\left[m_{k}\frac{\partial m_{i}}{\partial r_{j}}-m_{i}\frac{\partial m_{k}}{\partial r_{j}}\right]+\sum_{ijkl}C_{ijkl}\frac{\partial m_{i}}{\partial r_{j}}\frac{\partial m_{k}}{\partial r_{l}}\,.\label{K}\end{equation}
 The first term of Eq. (\ref{K}) is of special importance in the
phenomenological description of commensurability transitions \cite{tandt}.
As the free energy of a ferromagnet is invariant by the inversion
of the coordinate basis in the space, meaning $r_{i}\rightarrow-r_{i}$,
the antisymmetric term in $r_{i}$ must be also discarded. The actual
form of $K$ is ruled by the physical properties of the system and
by the point group symmetry of the OP. For a cubic crystal, the tensor
$C_{ijkl}$ is a number and the lowest order term is given by Eq.
(\ref{exchange}).

The second term of the free energy $f$ is the potential $U(m_{i})$,
described by a power series expansion in terms of the OP components,
\[
U\equiv\sum_{i_{1}}v_{i_{1}}^{(1)}m_{i_{1}}+\sum_{i_{1},i_{2}}v_{i_{1},i_{2}}^{(2)}m_{i_{1}}m_{i_{2}}+...+\sum_{i_{1},i_{2},...,i_{p}}v_{i_{1},i_{2},...,i_{p}}^{(p)}m_{i_{1}}m_{i_{2}}...m_{i_{p}}+O(m^{p+1}),\]
 provided that $m$ is small, where $i_{p}=1,...,n$ runs over the
components of the OP. The precise form of $U$ also depends on the
OP point group. In a ferromagnetic crystal, this term defines the
energy of crystalline anisotropy. The term of first order is clearly
excluded, due to the inversion symmetry of the crystal.

The ferromagnets can be classified according to their axes (or planes)
of easy magnetization. For crystals with one axis of anisotropy, say
along the {[}001{]} direction (or the $z$ axis), the second order
term is usually written in the form\[
v^{(2)}(m_{x}^{2}+m_{y}^{2})\quad\qquad\textrm{or}\qquad-v^{(2)}m_{z}^{2}.\]
 Both forms are equivalent, because they are related by an irrelevant
constant $v^{(2)}m_{0}^{2}$. If $v^{(2)}>0$ it is said that we have
an easy axis of magnetization, while for $v^{(2)}<0$ we have an easy
plane of magnetization, namely the $xy$ plane. The fourth order invariants
are made of free combinations of products between the two second order
invariants above. We see that there are four invariant terms of fourth
order entering in the free energy as a linear combination. The simplest example of 
uniaxial ferromagnet corresponds to a crystal with tetragonal symmetry. The hexagonal
lattice of cobalt is another example of uniaxial ferromagnet, with one
easy axis of magnetization perpendicular to the hexagonal lattice.
In fact, the anisotropy along the hexagonal directions is very small
and appears only in the sixth order terms \cite{Landau}. 
For biaxial crystals, we
need one more free parameter in the second order invariant of the
$xy$ plane,\[
v_{xx}^{(2)}m_{x}^{2}+v_{yy}^{(2)}m_{y}^{2}\,.\]
 In the particular case of isotropic cubic crystals, there are no
privileged directions between the three principal directions {[}100{]},
{[}010{]} and {[}001{]}. In this example, we have three equivalent
easy axes of magnetization. The lowest order term $U_{2}=\alpha\left[m_{x}^{2}+m_{y}^{2}+m_{z}^{2}\right]=\alpha m_{0}^{2}$
is spherically symmetric. In fourth order, the invariants can be written
in two different (and equivalent) ways \cite{Landau},

\[
U_{4}=-\frac{1}{2}v^{(4)}\left(m_{x}^{4}+m_{y}^{4}+m_{z}^{4}\right)\quad\textrm{or}\quad U_{4}=v^{(4)}\left(m_{x}^{2}m_{z}^{2}+m_{y}^{2}m_{x}^{2}+m_{y}^{2}m_{z}^{2}\right)\]
related by the constant $\frac{1}{2}v^{(4)}m_{0}^{4}$. 
After these examples, we conclude that we may set the general form of
the free energy by simple symmetry arguments only. To be more specific than this and specify the values of the parameters left, we need more information from experiments or from microscopic calculations. At the mean-field 
level, however, the qualitative form of the free energy is in general sufficient for drawing several important conclusions about the system.    

The last term of the free energy is due to the inclusion of the external
magnetic field $\mathbf{H}$\cite{hg}, \[
U_{B}\equiv-\mathbf{m}\cdot\mathbf{H}-\frac{1}{8\pi}H^{2}\,.\]
 The expression above satisfies the thermodynamic relation for the
total magnetic field $\mathbf{B}$ of the crystal,\[
\frac{1}{4\pi}\int_{V}\textrm{d}^{d}r\, B_{i}(\mathbf{r})=\frac{1}{4\pi}\int_{V}\textrm{d}^{d}r\left[H_{i}(\mathbf{r})+4\pi\langle m_{i}(\mathbf{r})\rangle\right]=-\frac{\partial G(\mathbf{m},\mathbf{H})}{\partial H_{i}}\,,\]
 where the Gibbs potential $G$ has been defined in Eq. (\ref{Z})$-$(\ref{action})
and $M_{i}=\int_{V}\textrm{d}^{d}r\,\langle m_{i}\rangle$ is the
total magnetization.

\section{Critical Region}

We will concentrate our focus in the simplest case, a crystal with
one axis of easy magnetization along the $z$ direction. Anisotropies
in the exchange are usually small, and (\ref{exchange}) is a good
approximation, even for axial symmetry. The dominant anisotropy effects
in the spin Hamiltonian come from the admixture of the spin-orbit
coupling into the crystal field. Similarly to (\ref{anisotropyTerm}),
we will write the second order contribution from the crystalline anisotropy
in the equivalent form\begin{equation}
\left[\alpha(m_{x}^{2}+m_{y}^{2})+\gamma m_{z}^{2}\right],\label{aniso2}\end{equation}
 with $\alpha$ and $\gamma$ positive constants. In the bulk, far
from the domain wall, we have

\[
m_{x}=m_{y}=0,\,\, m_{z}=\pm m_{0},\qquad\nabla m_{z}=0\,\,.\]
 To study the interface criticality, we write the excess free energy
relative to a bulk system in the following form:

\begin{eqnarray}
f(\mathbf{r},T) & = & a(T)\,\,\left[\alpha(m_{x}^{2}+m_{y}^{2})+\gamma\left(m_{z}^{2}-m_{0}^{2}\right)\right]+\nonumber \\
 &  & +b\,\left[\gamma^{2}\left(m_{z}^{4}-m_{0}^{4}\right)+2\alpha\gamma(m_{x}^{2}+m_{y}^{2})m_{z}^{2}+\alpha^{2}(m_{x}^{2}+m_{y}^{2})^{2}\right]+\label{free}\\
 &  & \qquad\qquad+c\,\left[|\nabla m_{x}|^{2}+|\nabla m_{y}|^{2}+|\nabla m_{z}|^{2}\right]\,\,.\nonumber \end{eqnarray}
 As usual, the phenomenological coefficient $a(T)\,$for the quadratic
term changes sign at the critical temperature and is taken as a linear
function of $T$, in the form \begin{equation}
a(T)=\left(\frac{T-T_{c}}{T_{c}}\right)a_{0}\equiv a_{0}\,\, t,\label{at}\end{equation}
 with $a_{0}>0$ and $t=\frac{T-T_{c}}{T_{c}}$, so $a(T)$ is negative
in the low temperature phase. We then assume $\alpha<\gamma$, since
the $z$-axis is taken as the easy axis of magnetization. Additionally,
we will assume that $\mathbf{H}\equiv0$.

The above Landau free energy (\ref{free}) is a simple generalization
of the commonly used expression to study phase transitions (see for
instance Ref.\cite{hg}, p. 417), with quadratic and quartic terms
in the order parameter. Due to the magnetic anisotropy, the elementary
invariants of the axial symmetry group are now $(m_{x}^{2}+m_{y}^{2})$
and $m_{z}^{2}$, and the free energy is built from them\cite{tandt}.
The fourth order term in the free energy can be interpreted as the
contribution of the spin quadrupole interaction. To minimize the number
of parameters, the quartic term is written as the square of the quadratic
one. So we are left with the set $(a(T),b,c,\alpha,\gamma)$ of free
parameters, where $\alpha$ and $\gamma$ are determined by the crystal
field of the ferromagnet. The others are usual Landau parameters,
with $b$ and $c$ slowly varying with temperature, even at the critical
point. We are assuming that they are constant. The behavior at large
$|\mathbf{m}|$ is dominated by the quartic term in (\ref{free}).
This requires the constant $b$ to be positive, in order to have minima
of the free energy at $\mathbf{m\neq}0$, for the low temperature
phase (where $a<0$). The constant $c$ is also positive, since it
costs some energy to create an interface.

The original treatment of Landau and Lifshitz argues that the minimization
of the energy has to be done in two different regions of the crystal:
\emph{i)} in the intermediate region between domains (interface),
where the contribution of the inhomogeneity cannot be neglected; and
\emph{ii)} in the region close to the surfaces, where the closed flux
configuration of the field requires appropriate boundary conditions.
We will use here the same argument, but for the more general free
energy given by (\ref{free}).

\subsection{Domain Wall Magnetization}

We shall concentrate first on the interface region. Without loss of
generality, we assume that the crystal is infinite and the interface
between domains is in the $yz$-plane. In the absence of a magnetic
field, due to symmetry, the spins are all in the $yz$-plane. Far
from the wall, they are aligned with the $z$-axis (parallel or anti-parallel).
Close to the walls, we assume that they are deflected by an angle
$\theta$ from the$\,\, z$-axis, with components:

\begin{equation}
m_{x}=0,\qquad m_{y}=m_{0}\sin\theta,\qquad m_{z}=m_{0}\cos\theta,\label{components}\end{equation}
 where $\theta\equiv\theta(x)$ is a function of $x$ only. Note that
$m_{0}=\left|\mathbf{m}\right|$ is temperature dependent, but considered
uniform in space. The substitution of (\ref{components}) into (\ref{free})
leads to \begin{equation}
\begin{array}{c}
f\,\,\left[\theta(x),\theta^{\prime}(x);T\right]=cm_{0}^{2}(\theta^{\prime})^{2}+a(T)[\alpha\!\!+\!\!(\gamma-\alpha)\cos^{2}\theta]m_{0}^{2}-a(T)\gamma m_{0}^{2}+\\
\\\qquad\qquad\qquad\qquad+b\,[\gamma^{2}\cos^{4}\theta+2\gamma\alpha\sin^{2}\theta\cos^{2}\theta+\alpha^{2}\sin^{4}\theta]m_{0}^{4}-b\gamma^{2}m_{0}^{4}.\end{array}\label{free_theta}\end{equation}
 In order to apply the Euler variational principle to the action (\ref{action}),
one notes that the `Lagrangian ' $f\,\,\left[\theta(x),\theta^{\prime}(x);T\right]$
does not explicitly depend on the variable $x$. In this case, the
Euler-Lagrange equation can be written as \[
\frac{d}{dx}\left[\theta^{\prime}(x)\frac{\partial f}{\partial\theta^{\prime}}-f\right]=0\,\,,\]
 which reduces to

\begin{eqnarray}
cm_{0}^{2}(\theta^{\prime})^{2} & = & K+a(T)[\alpha\!\!+\!\!(\gamma-\alpha)\cos^{2}\theta]m_{0}^{2}-a(T)\gamma m_{0}^{2}+\nonumber \\
\label{euler}\\ &  & +b\,[\gamma^{2}\cos^{4}\theta+2\gamma\alpha\sin^{2}\theta\cos^{2}\theta+\alpha^{2}\sin^{4}\theta]m_{0}^{4}-b\gamma^{2}m_{0}^{4}.\nonumber \end{eqnarray}
 where $K$ is a constant to be evaluated using the boundary conditions
(BC). Far from the interface between domains, we set the spins asymptotically
aligned with the $z$-axis, remembering that $z$ is our easy magnetization
direction. To satisfy the closure configuration, we impose

\begin{equation}
\theta=\left\{ \begin{array}{l}
0,\,\, x\rightarrow-\infty\,\,,\\
\\\pi,\,\, x\rightarrow\infty\,\,,\end{array}\right.\label{bc1}\end{equation}
 and

\begin{equation}
\theta^{\prime}=0\,\,,\rightarrow\pm\infty\,\,,\label{bc2}\end{equation}
 which result in $K=0$. The equation to be solved now is

\begin{equation}
cm_{0}^{2}(\theta^{\prime})^{2}=a(T)[(\alpha-\gamma)\sin^{2}\theta]m_{0}^{2}+bm_{0}^{4}\left[2\gamma\alpha\sin^{2}\theta\cos^{2}\theta+\alpha^{2}\sin^{4}\theta+\gamma^{2}(\cos^{4}\theta-1)\right]\,\,,\label{balance}\end{equation}
 which means that the free energy is minimized when the exchange energy
density is equal to the anisotropy one. Equation (\ref{balance})
can be written in the form

\begin{equation}
(\theta^{\prime})^{2}=\sin^{2}\!\theta\,\,(A+B\cos^{2}\theta)\,\,,\label{reduced}\end{equation}
 whose solution can be given as

\begin{equation}
\frac{\sqrt{1+\frac{B}{A}}\cos\theta}{\sqrt{1+\frac{B}{A}\cos^{2}\theta}}=-\tanh[x\sqrt{A+B}\,\,]\,\,,\label{solu}\end{equation}
 where $A$ and $B$ are

\begin{eqnarray}
A & = & \frac{a_{(T)}}{c}(\alpha-\gamma)\;\left[1+\frac{b}{a_{(T)}}m^{2}(\alpha+\gamma)\right]\nonumber \\
\label{coeff}\\B & = & \frac{a_{(T)}}{c}(\alpha-\gamma)\;\left[-\frac{b}{a_{(T)}}m^{2}(\alpha-\gamma)\right]\,\,.\nonumber \end{eqnarray}

The above equations give the distribution of the magnetic moment density
in a general crystal between two neighboring domains. There are two
equivalent solutions with opposite helicities. The width of the interface
(Bloch wall) is given by \begin{equation}
\lambda=\frac{1}{\sqrt{A+B}}.\label{bloch}\end{equation}

To check the consistency of (\ref{solu}), we remind the reader that
$a(T)$ is a negative scalar function for $T<T_{c}$. Noting that
$\alpha<\gamma$ (easy $z-$direction), the condition $\left(A+B\right)>0$,
will impose limitations for the lower bound of $\alpha$. In our discussion,
we are considering the limit of small anisotropy, for which $\alpha\lesssim\gamma$.
As we will show below, mean field implies that $bm_{0}^{2}/a(T)$
is finite at the critical point, which determines the critical exponent
of the magnetization. We will discuss those points later on.

The symmetry of the magnetization planes in the presence of an uniaxial
anisotropy is represented by the $C_{2v}$ point group, which admits
two invariant terms, $m_{0}^{2}$ and $m_{0}^{2}\cos2\theta$ in the
expansion of the free energy \cite{tandt}. This is easily noticed
by decomposing the anisotropy term $U_{2}$, $m_{0}^{2}\left(\alpha\sin^{2}\theta+\gamma\cos^{2}\theta\right)$
{[}see (\ref{aniso2}) and (\ref{components}){]}, into the polynomial
basis formed by the two invariants, giving\[
\frac{1}{2}\left[(\alpha+\gamma)m_{0}^{2}+(\gamma-\alpha)m_{0}^{2}\cos(2\theta)\right].\]
 Clearly, the full symmetry term $m_{0}^{2}$ due to the paramagnetic
phase ($O(2)$ point group) competes with the other term representing
the symmetry of the domain ordered phase ($C_{2v}$ point group).
The stability of the domain walls is ruled by the anisotropy parameter
$\varepsilon$, whose proper expression is\[
0<\varepsilon\equiv\frac{\gamma-\alpha}{\gamma+\alpha}<1\,.\]
 Assuming that the anisotropy is small ($\varepsilon\ll1$), the ratio
$\frac{B}{A}$ is also small provided that $\left(A+B\right)>0$,
\[
\frac{B}{A}\approx\frac{2\alpha bm_{0}^{2}}{a\left(T\right)}\cdot\frac{\varepsilon}{\left[1+\frac{2\alpha bm_{0}^{2}}{a\left(T\right)}\right]}\,\,,\]
 and the solution can be expanded in terms of $\frac{B}{A}$

\begin{equation}
\left(1+\frac{B}{2A}\right)\cos\theta\left[1-\frac{B}{2A}\cos^{2}\theta\right]=-\tanh[\sqrt{A}\,\,(1+\frac{B}{2A})\,\, x]\,\,.\label{solt}\end{equation}
 Fig. 1 shows the distribution of the magnetization angle $\theta$
through the interface. The quantity $\left(A+B\right)$ vanishes at
the critical point yielding the limit $\lambda\rightarrow\infty$
and the solutions $\theta=\pm\frac{\pi}{2}$ everywhere. At first
sight, this solution may seem inconsistent with the boundary condition
imposed. However, we observe that the magnetization intensity goes
to zero at $T_{c}$. In any case, this behavior signals that
something odd is happening at the critical point and questions the
validity of mean field solutions there. The limit $\lambda\rightarrow\infty$
means that magnetic fluctuations are paramount at $T_{c}$.

%

\begin{figure}[htbp]
\centering
\includegraphics[width=0.55\textwidth]{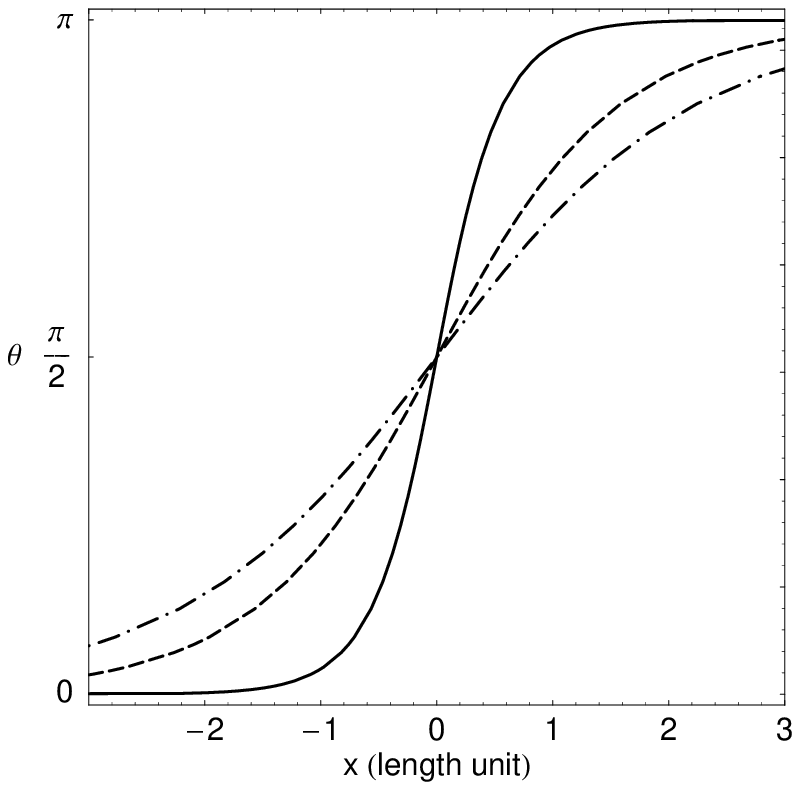}
\caption{Magnetization profile in the vicinity of a domain wall. The
solid curve represents the solution in the saturated regime at low
temperature (asymptotic case for $B\rightarrow0$ and $A=8$, now
normalized to adimensional parameters). The dashed and dot-dashed
curves were obtained solving (\ref{solt}), keeping the anisotropy
fixed and varying the temperature in direction to $T_{c}$, where
both, $A$ and $B$ vanish (dashed for $B=0.1,\,\, A=1$and dot-dashed
for $B=0.05,\,\, A=0.5$, respectively).}
\label{fig1}
\end{figure}

\subsection{Surface Energy}

Next we consider a finite crystal, where the domain structure is organized
in layers, as shown in Fig.2. We already know the spin distribution
through the Bloch wall, and want to calculate the width of magnetic
domains in a finite volume. This is done in a variational way, as
in the original contribution by Landau and Lifshitz\cite{ld}. We
note that the flux closure condition induces the formation of small
domains near the surfaces, where the magnetization points perpendicular
to the easy direction.

We proceed to the calculation of the wall energy. If $l$, $l_{x}$
and $l_{y}$ are the dimensions of the crystal in the $z$, $x$ and
$y$ directions respectively, the energy associated with one interface
(wall) between two domains is

\begin{eqnarray}
E_{wall} & = & ll_{y}\int_{-\infty}^{\infty}\! dx\,\, f(\mathbf{m},\mathbf{T})=\nonumber \\
 & = & ll_{y}\int_{-\infty}^{\infty}\! dx\,\,\left[cm_{0}^{2}(\theta^{\prime})^{2}+a(T)[\alpha\!\!+\!\!(\gamma-\alpha)\cos^{2}\theta]m_{0}^{2}-a(T)\gamma m_{0}^{2}+\right.\nonumber \\
 &  & \qquad\left.+b\,[\gamma^{2}\cos^{4}\theta+2\gamma\alpha\sin^{2}\theta\cos^{2}\theta+\alpha^{2}\sin^{4}\theta]m_{0}^{4}-b\gamma^{2}m_{0}^{4}\right]\label{wall}\end{eqnarray}
 where the limits of integration have been extended to $\left]-\infty,\infty\right[$
, considering that the layer width is much larger that the wall region.
The dominant contribution to this integral is concentrated inside
domain walls, where $\theta$ is close to $\pi/2$. This way, we may
neglect the cubic term in $\cos(\theta)$ in equation (\ref{solt}).
Substituting expressions (\ref{balance}) into (\ref{wall}) yields
an expression only in terms of the anisotropy \begin{eqnarray*}
E_{wall} & = & 2ll_{y}\left(\alpha-\gamma\right)\int_{-\infty}^{\infty}\! dx\,\,\sin^{2}\!\theta\left[a(T)m_{0}^{2}+\right.\\
 &  & \qquad\left.+b\,[\left(\alpha+\gamma\right)-\left(\alpha-\gamma\right)\cos^{2}\theta]m_{0}^{4}\right]\,\,,\end{eqnarray*}
 where we have used the form given by (\ref{reduced}). To lowest
order in the anisotropy $\varepsilon$ we get the result \begin{eqnarray*}
E_{wall} & = & -4ll_{y}\alpha m_{0}^{2}\,\,\varepsilon\,\,\left[a(T)+2b\alpha m_{0}^{2}\right]\int_{-\infty}^{\infty}\!\frac{dx}{\cosh^{2}\left(x\sqrt{A}\right)}\,\,\\
 & = & 4ll_{y}m_{0}^{2}c\sqrt{A}\,\,.\end{eqnarray*}
 It then follows that the total energy of $\left(l_{x}/\! d\right)$
walls is

\[
E_{1}=4\frac{ll_{x}l_{y}}{d}c\,\, m_{0}^{2}\sqrt{A}\,\,,\]
 where $d$ is the layer width.

Now, we turn to the surfaces for the flux closure configuration. The
ferromagnetic phase of the body produces some amount of field given
by the Maxwell equation $\nabla\cdot(\mathbf{H}+4\pi\mathbf{m})=0$
inside the sample, and $\nabla\cdot\mathbf{H}=0$ outside. Near the
surface and far from domain walls, the exchange contribution to the
energy is zero. The closed flux configuration, with no field lines
outside the sample, is favored when the anisotropy is small. Under
the above conditions, we get \[
\mathbf{H=0,\quad}\mathrm{and}\quad\nabla\cdot\mathbf{m}=0\,\,,\]
 what means that the spin distribution satisfies the boundary condition
$\mathbf{m}\cdot\mathbf{n}=0$, with $\mathbf{n}$ normal to the surface.
This way, surface poles are avoided and the global spin field inside
the crystal has no singular points as shown in Fig.2. For surface
domains, the main contribution to the energy density comes from the
anisotropy. At the surface, we get \[
m_{x}=\pm m_{0},\qquad m_{y}=0\,\,,\qquad m_{z}=0\,\,,\]
 and going back to (\ref{free}) we get the energy density

\[
f_{surface}=a(T)m_{0}^{2}(\alpha-\gamma)\,\left[1+\frac{bm_{0}^{2}}{a(T)}\left(\alpha+\gamma\right)\right]\,\,,\]
 which is again proportional to the anisotropy $\varepsilon$. For
a finite sample, we have two opposite surfaces at $z=0$ and $z=-l$.
Associating the volume $\left(l_{y}d^{2}/4\right)$ to a single surface
domain, and summing over $\left(2l_{x}/\! d\right)$ of such domains,
the total surface energy is given by \[
E_{2}=\frac{1}{2}l_{x}l_{y}d\,\, m_{0}^{2}c\,\, A\,\,.\]

\begin{figure}
\begin{center}\includegraphics[%
  scale=0.28]{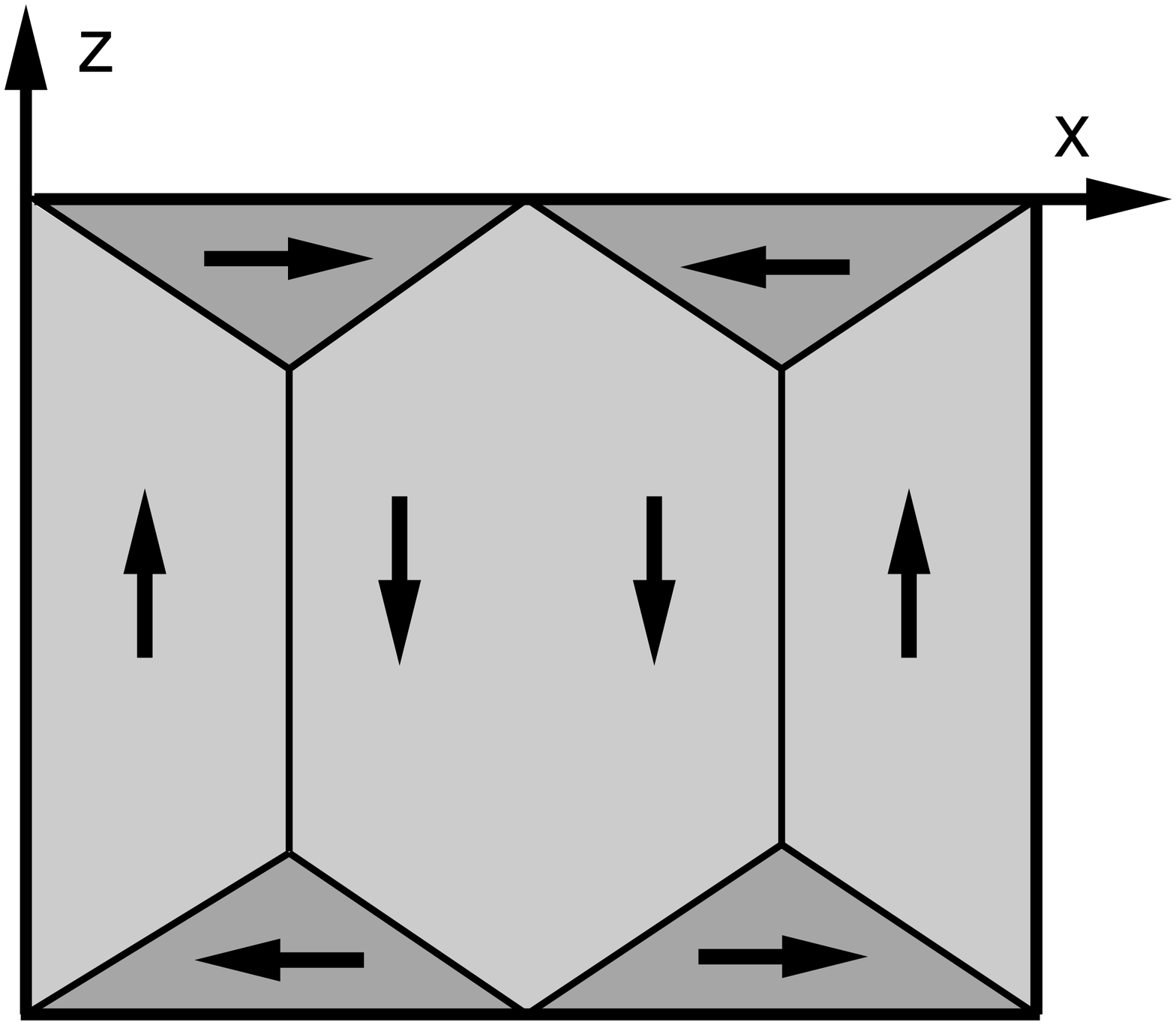}\end{center}

\caption{{\small Magnetic permeability distribution in the crystal for flux
closure. }}
\end{figure}

We then minimize the total energy $E=E_{1}+E_{2}$ simultaneously
in relation to $d$ and $m_{0}$\[
\left\{ \begin{array}{l}
\frac{\partial E}{\partial d}=0\,\,,\\
\\\frac{\partial E}{\partial m_{0}}=0\,\,,\end{array}\right.\]
 yielding the results

\begin{equation}
\left\{ \begin{array}{l}
d=2\sqrt{2l}\!\!\,\,\sqrt{\sqrt{\frac{1}{A}}}\,\,,\\
\\\frac{b\, m_{0}^{2}}{a(T)}=-\frac{4}{7}\frac{1}{(\alpha+\gamma)}\quad.\end{array}\right.\label{variation}\end{equation}

The quantity $A$, defined in (\ref{coeff}), vanishes at the critical
point, thus causing the divergence of the domain width at $T_{c}$.
The associated critical exponent is different from the one that gives
the divergence of the wall width $\lambda$ in (\ref{bloch}) \[
\lambda^{-1}=\sqrt{A+B}=\sqrt{\left(\frac{\alpha-\gamma}{c}\right)a(T)\left[1+\frac{2\gamma bm_{0}^{2}}{a(T)}\right]}\quad.\]
 In the expression above we encounter the quantity \[
D\equiv1+\frac{2\gamma bm_{0}^{2}}{a(T)}\,\,,\]
 that has to be positive in order to get real solutions. From (\ref{variation}),
we see that this is the case for $\gamma/7<\alpha<\gamma$. So, we
are on safe grounds for small anisotropy, $\alpha\lesssim\gamma$.
We note that the critical exponent for the magnetization is the same
as in the homogeneous case ($\beta=1/2$), but the full expression
is different from the one obtained in conventional mean field theory.
It is worth to note that the isotropic, homogeneous case, satisfies
$D=0$, \emph{i.e.} no domains are present.

From (\ref{variation}),
we get the result \[
m_{0}=\left\{ \begin{array}{ll}
0 & \mbox{, $T>T_{c}$}\\
\sqrt{-\frac{4a(T)}{7b\,(\alpha+\gamma)}} & \mbox{, $T<T_{c}$}\,\,.\end{array}\right.\]
The two scales, $d$ and $\lambda$, diverge at the critical point,
but the divergence of the wall width is faster (exponent $-1/2$)
than the one for the domain width (exponent $-1/4$), thus showing
that the critical point marks the onset of strong magnetic fluctuations.

For magnetic thin films,
the Bloch walls produce magnetostatic poles in the surface of the
film, enhancing enormously the energy of the crystal when the thickness
is smaller than $\sim10^{-6}$cm \cite{chikasumi}.
In general, the crystal prefers to deflect the spins of the 
wall along the directions parallel to the thin film surface, giving rise to
a Ne\'el wall, when the easy magnetization direction is in the plane of the 
film, as shown schematically in Fig. 3.

Despite it is tempting to generalize our previous conclusions to this system, 
the long-range dipolar interactions (which we have neglected so far) are relevant in an infinite slab of small thickness,
not mattering how weak the dipolar interaction is in comparison to the
nearest neighbor exchange interaction \cite{garel}.
To illustrate how important this interaction might be, in some ferromagnets, the 
dipolar interactions are responsible for the appearance of domains 
in the complete absence of crystalline field effects, when the free energy is fully isotropic 
in a given plane of magnetization, for example. In these systems, 
the exchange competes with the dipolar interactions, stabilizing a non-homogeneous phase with 
a finite wave-vector in an arbitrary direction of the magnetization plane. Because of the large 
phase space of all the degenerate wave-vector directions allowed by symmetry, the fluctuations are 
strongly enhanced at $T_c $, giving rise to the fluctuation induced first order transitions 
studied by Brazovskii \cite{brazovskii}. In this specific case, the study of the phase transition is totally out of the scope of the mean field analysis. Here, we have focused our study in ordinary 3D ferromagnets, where the 
dipolar contribution is not a relevant interaction.

\begin{figure}[htbp]
\centering
\includegraphics[width=0.6\textwidth]{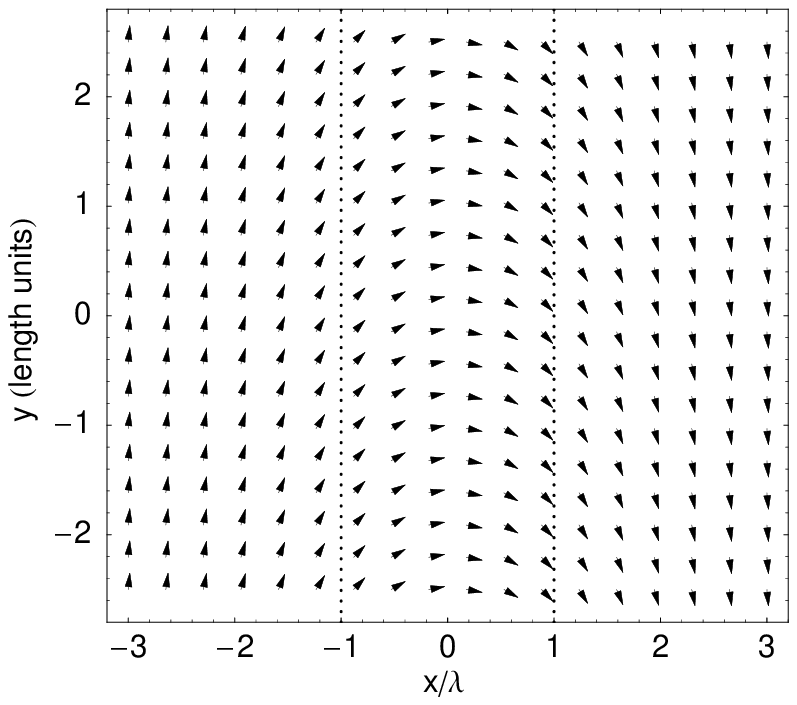}
\caption{Illustration of a 180$^{\textrm{o}}$ Ne\'el wall.}
\label{fig2}
\end{figure}

\subsection{Critical Exponents}

When temperature approaches $T_{c}$ from below, $a(T)$ as defined
in (\ref{at}), goes to zero linearly, and the entire system is affected
by long range fluctuations. At $T_{c}$, the correlation length diverges
and the system is scale invariant\cite{hg}. In our model, this is
signaled by the divergence of the wall width $\lambda$, which plays
the role of the length scale and goes to infinity under the power
law $\lambda\propto|t|^{-\frac{1}{2}}$. Since the domain width behaves
as $d\propto|t|^{-\frac{1}{4}}$, we find that the wall region enlarges
and `compresses' the domains as temperature raises in direction to
$T_{c}$, meaning that domains firstly lose their identity and finally
disappear at the critical point. We observe that all the critical
exponents in our treatment refer to behaviors below $T_{c}$ (in critical
phenomena, one distinguishes the behaviors above and below $T_{c}$).

In what follows, we advance conclusions derived from the previous
calculation. To calculate the heat capacity, we compute the total
internal energy of the system as

\[
E=2c\sqrt{2l}\, l_{x}l_{y}\, m_{0}^{2}A^{\frac{3}{4}},\]
 once the minimization process (\ref{variation}) is done. Note that
this is the excess energy relative to the bulk system, associated
with surfaces and interfaces. From this, we calculate the specific
heat $C=T\frac{\partial^{2}E}{\partial T^{2}},$ and find that $C\propto|t|^{-\alpha}$
diverges at $t=0$, with the critical exponent $\alpha=1/4$, which
is different from the standard mean field approximation $\alpha_{B}=0$
for homogeneous systems (bulk magnetization) and also different from
standard mean field results for interfaces ($\alpha_{S}=1/2$)\cite{binder}.
The different critical behavior comes here from the criticality of
the domain width $d$.

We also know from experiments that the static magnetic susceptibility
diverges at the critical point. In our calculation we get \[
\chi=\left(\frac{\partial^{2}E}{\partial m_{0}^{2}}\right)^{-1}\propto|t|^{-\frac{3}{4}},\,\,\mathrm{for\,\,}t\ll1\,\,,\]
 and therefore $\gamma=\frac{3}{4}$, which also disagrees with the
standard mean field value $\gamma=1$. Applying the Josephson, Fisher
and Widom scaling laws\cite{hg,st}\[
\nu d=2-\alpha\qquad\textrm{and}\qquad\gamma=\nu(2-\eta)=\beta(\delta-1)\]
 for the correlation length $\xi\propto t^{-\nu}$, for the equation
of state $M=\int_{V}\textrm{d}^{d}x\,\langle m\rangle\propto H^{-\frac{1}{\delta}}$,
and for the correlation function $(\Gamma)$ power law $p=d-2+\eta$,
with\[
\Gamma\propto r^{-p}\textrm{e}^{-r/\xi}\,,\]
 we find that $\nu=\frac{7}{12}$, $\eta=\frac{5}{7}$ and $\delta=\frac{5}{2}$. 

The different exponents and the singularity in the heat capacity are
explained by the finite size effects manifested in the geometry of
the ferromagnet surfaces, which lead to the formation of Bloch walls
between magnetic domains. This statement can be easily tested by removing
the closed flux boundary conditions in the thermodynamic limit. In
this limit, the magnetization will be homogeneously distributed as
if the crystal had a single domain with the size of the system, \textit{i.e.} $d\sim l_{x}$
not mattering the temperature. We regain all the standard mean field
exponents in this situation.

\section{Mesoscopic Systems}

The main effect resulting from the introduction of an external magnetic
field $\mathbf{H}$ in the ferromagnet is the displacement of the
domain walls from the zero field equilibrium position. As the bulk
energy of each domain is enhanced or reduced by the magnetic coupling
of the magnetization with the magnetic field,\[
-\mathbf{m}\cdot\mathbf{H}\,,\]
 depending on the orientation of the spins with respect to $\mathbf{H}$,
the domain walls are dislocated in the direction that reduces the
size of energetically unfavorable domains, enlarging the favorable
ones and reestablishing the equilibrium. As shown by Landau and Lifshitz
\cite{ld}, the dynamical interaction of the magnetization with the
magnetic field is driven by the Zeeman effect, where the spins start
to precess around $\mathbf{H}$ as free moments, and by the relativistic
interaction (with coupling $\eta$), following the equation of motion\[
\stackrel{\cdot}{\mathbf{m}}\,=\mu_{0}\left[\mathbf{H}\times\mathbf{m}+\eta\left(\mathbf{H}-(\mathbf{H}\cdot\mathbf{m})\frac{\mathbf{m}}{m_{0}^{2}}\right)\right],\]
 where $\mu_{0}=e/mc$, with $e$ the electron charge, $m$ the electronic
mass and $c$ the speed of light. The dot represents a time derivative.
Considering the case where $\mathbf{H}$ is oriented along the easy
magnetization direction, they found the domain walls speed of displacement
to be\[
v\cong\frac{\mu_{0}m_{0}^{2}}{\eta}H\sqrt{\left(\frac{\alpha-\gamma}{c}\right)a}\,,\]
 employing the notation of sec. 2 for the saturated regime ($a=const.$
and $b=0$). If the magnetic field is constant in time, the velocity
will be reduced to zero by the magnetic pressure of the bulk in a
second moment, as the system approaches the equilibrium. On the other
hand, if we orient the external magnetic field transversely to the
easy magnetization axis, the domain walls will remain in equilibrium
with the magnetization of the bulk. The application of external magnetic
fields on domain walls may have important applications in the fabrication
of switches for bulk spin polarized electrons in mesoscopic systems. 

The scattering of an electron through a 180$^{\textrm{o}}$ domain
wall barrier comprehends two cases of technological interest for the
fabrication of switches. Consider for example one electron of spin
up, aligned with the magnetization of the bulk, flowing in direction
to a domain wall. If the time of flight of the electron through the
wall is long enough, its spin will be adiabatically deflected and
the scattered electron will be transmitted to the other side of the
barrier with the spin flipped down, causing no additional cost of
energy to the bulk. On the other hand, if the time of flight is too
short, (considering that the electron is too fast or the barrier is
too narrow) the spin will not have time enough to be deflected by
the spiral of magnetization and the electron (in the classical picture)
will be reflected back from the barrier, in order to save energy from
the bulk.

As shown by Cabrera and Falicov \cite{cabrera}\emph{,} the domain
walls may respond for a large change in the resistivity of metallic
ferromagnets when the difference in the density of states
between majority and minority spins at the Fermi surface is also large. 
Consider that we
have two spherical Fermi surfaces of different sizes, for example,
one for spin up electrons and another for spin down ones. In the situation
where the radius of one of the bulk Fermi surfaces is very small (minority
spin) and the other very large (majority one), the majority spin electrons
in one side of the barrier will find few channels to tunnel ballistically
to the other side, where they would occupy the minority spin states.
This way, if the wall is sufficiently narrow in comparison to the
electron mean free path, the electrons will most probably be backscattered
by the barrier. The change in the bulk resistivity predicted, however,
is not so large as to induce giant and colossal magnetoresistances.
In contrast, the magnetoresistance can grow several orders of
magnitude when the width of the walls becomes small in comparison
to its size in bulk, as it has been observed experimentally in
magnetic nanojunctions.

A very interesting physics shows up in mesoscopic systems when a domain
wall is geometrically constrained by a small constriction separating
two ferromagnetic bulks of wider cross section. Bruno \cite{bruno}
derived the somehow remarkable result that when the constriction cross
section is much smaller than the bulk one, the domain walls are practically
independent on the specific characteristics of the material, including
crystalline anisotropy and exchange stiffness, and depend only on
the geometry of the constriction. The free energy of the system is
in the form\begin{equation}
f\left[\theta(x),\theta^{\prime}(x)\right]=\left[cm_{0}^{2}(\theta^{\prime})^{2}+U(\theta)\right]S(x)\,,\label{freeH2}\end{equation}
 where $S(x)$ defines the geometry of the constriction and $U$ is
the crystalline anisotropy term. Proceeding with the minimization,
one finds through the usual Euler-Lagrange equation \[
\frac{\textrm{d}}{\textrm{d}x}\frac{\partial f}{\partial\theta^{\prime}}-\frac{\partial f}{\partial\theta}=0\]
 that\[
\theta^{\prime\prime}+\theta^{\prime}\frac{S^{\prime}}{S}-\frac{1}{2cm_{0}^{2}}\frac{\partial U}{\partial\theta}=0\,.\]
 Bruno proposes that if the second term is much larger than the third
and in addition if $S$ is integrable, meaning if $\int_{-\infty}^{\infty}S(x)\textrm{d}x$
is finite, then we may drop the crystalline contribution. Defining
the wall width according to the criterion\[
\lambda=4\left[\int_{-\infty}^{\infty}\textrm{d}x\,(\theta^{\prime})^{2}\right]^{-1}=4\left[\int_{-\pi/2}^{\pi/2}\textrm{d}\theta\,\theta^{\prime}\right]^{-1},\]
 he finds the simple results:\[
\theta(x)=\pi\frac{\int_{-\infty}^{x}S^{-1}(u)\textrm{d}u}{\int_{-\infty}^{\infty}S^{-1}(x)\textrm{d}x}-\frac{\pi}{2}\,,\]
 and\[
\lambda=\frac{4}{\pi^{2}}\frac{\left[\int_{-\infty}^{\infty}\textrm{d}x\, S^{-1}(x)\right]^{2}}{\int_{-\infty}^{\infty}\textrm{d}x\, S^{-2}(x)}\,,\]
 which are defined by the constriction only. In the most general case,
where the crystalline term $U$ is present, he finds for different
models of constriction that the domain wall is still driven by $S$,
and not by $U$ or the exchange stiffness $c$. 

When a magnetic field is applied along the easy magnetization axis,
the domain wall inside the constriction suffers a finite displacement
proportional to the applied field, where the new position of equilibrium
is achieved thanks to the magnetostatic pressure played by the constriction
geometry \cite{garcia}. The localization of domain walls in nanojuctions
are a theoretical possibility for the fabrication of fast switches
demanded for \emph{spintronic} devices.

In a wide sense, the word spintronics means manipulation and control
of the spin degrees of freedom in condensed matter systems, and to
use this knowledge to make useful devices \cite{zutic}. Here, we
focus on spin transport properties in ferromagnetic metals. We have
already discussed the magnetoresistance effect (MR) due to the domain
structure and the scattering of electrons from Bloch walls. The resistance
of a ferromagnetic sample changes with the application of a small
magnetic field. The field reorients the domains, and for a large enough
value, it saturates the sample. The saturated configuration has a
smaller resistance (negative magnetoresistance) than the one with
domains, where the magnetization is not homogeneous (for a $180^{o}$
wall, the magnetization of one domain is antiparallel to the magnetization
of the adjacent domain)\cite{cabrera}. This simple system suggests
the invention of {}``artificial'' devices where one could monitor
the magnetic configuration through the application of small magnetic
fields. Based on the same physics, magnetic tunneling junctions (MTJ)\cite{mtj}
and metallic multilayers (MML) structures\cite{mml} were conceived,
in order to operate as \emph{spin valves}, due to the `giant' magnetoresistance
(GMR)effect displayed by those systems.%
\footnote{The MR effect is called GMR for values exceding 10\%, which are one
order of magnitude larger than the typical values of the anisotropic
magnetoresistance used commercially. MTJ prototypes currently produced
nowadays present a GMR in the range 25\%-30\%
at room temperature. In MML systems, the MR may be as large as 50\%
- 65\% , but saturating fields are substantially higher
(about 1 T in the original system of Ref. \cite{mml}).%
} In a typical MTJ, two
ferromag\-netic metallic electrodes are separated by a narrow nonmagnetic
insulating layer. For MML, two or more ferromagnetic layers are separated
by nonmagnetic metallic spacers. In both structures, one can pin the
magnetization of one of the electrodes using an additional magnetic
layer which is strongly exchanged-biased with the ferromagnet\cite{schuller}.
The other ferromagnetic electrode is left `free' to orient its magnetization
with small applied fields. One can then change the resistance of the
device by manipulating the relative orientation of the magnetization
of both electrodes, phenomenon that has been named as spin-valve effect
and is currently used in commercial applications for magnetic recording\cite{falicov,parkin}.
For technological applications, ideal devices should have a GMR as
large as possible, obtained at room temperature with small applied
fields. Experimental measurements of GMR in magnetic multilayers are
usually performed for the current-in-plane (CIP) geometry, where the
electric current flows parallel to the layers. For this setup, the
resistance is fairly large and can be measured using standard techniques.
The other possibility, much more complicated for experimental implementation,
is to do the measurement with the current perpendicular to the layer
plane, the so called CPP geometry (CPP stands for current-perpendicular-to-the-plane).
In this case, the resistance is very low and can only be measured
using extremely sensitive techniques\cite{gijs}.

Domain walls are thought to play a relevant role in recent GMR experiments
in magnetic nanocontacts. Large values of MR, of the order of 300\%-3000\%
at room temperature and for fields of about 100 Oe, have been reported
in the literature\cite{nico,chopra}. The MR values found are attributed
to strong electron scattering from narrow domain walls which are formed
in the contact region. Due to the constricted geometry and the rapid
variation of the magnetization across the domain wall, the electron
spin cannot follow adiabatically the local magnetization, as it is
the case in bulk ferromagnets\cite{cabrera}.

\section{Conclusion}

We have developed an illustrative application of the Landau mean
field theory for phase transitions in ferromagnetic bodies. The advantage
of this procedure resides in the phenomenological nature of the free
energy parameters, allowing us to calculate equations of state without
stating a specific microscopic model. The classical Landau approach
can then be extended to describe the formation of domains, with the
corresponding change of the critical behavior, still within the mean
field approximation. Despite this theory is in fact very crude to
describe the neighborhood of the critical point for systems whose
specific heat diverges at the phase transition \cite{st}, the theory 
is also known for making
remarkable qualitative predictions, with deep insights on the physics 
of critical phenomena, even in situations that 
extrapolate its region of validity\cite{kadanoff}. In particular, the 
theory provides its own criterion of failure and
suggests new procedures to correctly describe the physics at the critical
point, where fluctuations must be included\cite{birgenau}.

We have shown that the problem of domains in ferromagnets is driven by 
the presence of two different length
scales competing in the vicinity of the critical point. On one side,
there is the bulk correlation length which drives the system to a
classical mean field ferromagnetic behavior, and on the other, the
length scale of the interfaces between domains. The competition between
them produces two major effects: \textit{i)} a new critical behavior, where
the specific heat is strongly enhanced near $T_{c}$ from below in
comparison to the bulk homogeneous limit, but is still small in comparison to the
limit of surfaces, giving an intermediary critical exponent $\alpha_{B}<\alpha<\alpha_{S}$;
and \textit{ii)} a raise in the magnetoresistance in the ordered phase,
specially for ferromagnets with a layered domain wall geometry. 
Despite the study of magnetic domains is a rather old subject, it
has recently opened  a very promising field in technology. The transport
properties of ferromagnetic domains suggest several interesting applications
in spintronic devices, which we have briefly discussed.

\subsection*{Acknowledgments}

The authors acknowledge partial support from \emph{Funda\c{c}\~{a}o
de Amparo \`{a} Pesquisa do Estado de S\~{a}o Paulo} (FAPESP, Brazil)
through the project 98/01289-2 and \emph{Conselho Nacional de Desenvolvimento
Cient\'{\i}fico e Tecnol\'{o}gico} (CNPq, Brazil), project 301221/77-4/FA.


\end{document}